\documentclass[preprint,12pt,numbers]{elsarticle}

\usepackage{amssymb}
\usepackage{float}
\usepackage{booktabs}
\usepackage{tikz}
\usetikzlibrary{positioning,shapes,arrows.meta}
\usepackage{booktabs,tabularx,ragged2e,array}
\newcolumntype{L}{>{\RaggedRight\arraybackslash}X}
\usepackage{amsmath, amssymb}
\usepackage{graphicx}
\usepackage{enumitem}
\usepackage{hyperref}
\usepackage{titlesec}

\usepackage{url} 

\usepackage{xcolor}

\usepackage{pgfplots}
\pgfplotsset{compat=1.18}

\usepackage{csvsimple}  
\usepackage{ifthen}     
\usepackage{siunitx}    

\usepackage{amsmath}

\begin{document}

\makeatletter
\def\ps@pprintTitle{%
   \let\@oddhead\@empty
   \let\@evenhead\@empty
   \let\@oddfoot\@empty
   \let\@evenfoot\@oddfoot
}
\makeatother

\begin{frontmatter}


\title{Full-Stack Knowledge Graph and LLM Framework for Post-Quantum Cyber Readiness}
\author[cci]{\corref{cor1}Rasmus Erlemann}

\ead{rerleman@charlotte.edu}

\cortext[cor1]{Corresponding author.}

\author[cci]{Charles Colyer Morris}
\author[cci]{Sanjyot Sathe}

\affiliation[cci]{organization={College of Computing and Informatics, University of North Carolina at Charlotte},
            addressline={9201 University City Blvd}, city={Charlotte}, state={North Carolina},
            postcode={28223}, country={United States of America}}

\begin{abstract}
The emergence of large-scale quantum computing threatens widely deployed public-key cryptographic systems, creating an urgent need for enterprise-level methods to assess post-quantum (PQ) readiness. While PQ standards are under development, organizations lack scalable and quantitative frameworks for measuring cryptographic exposure and prioritizing migration across complex infrastructures. This paper presents a knowledge graph based framework that models enterprise cryptographic assets, dependencies, and vulnerabilities to compute a unified PQ readiness score. Infrastructure components, cryptographic primitives, certificates, and services are represented as a heterogeneous graph, enabling explicit modeling of dependency-driven risk propagation. PQ exposure is quantified using graph-theoretic risk functionals and attributed across cryptographic domains via Shapley value decomposition. To support scalability and data quality, the framework integrates large language models with human-in-the-loop validation for asset classification and risk attribution. The resulting approach produces explainable, normalized readiness metrics that support continuous monitoring, comparative analysis, and remediation prioritization.
\end{abstract}

\tnotetext[t1]{The authors gratefully acknowledge Giacinto Paulo Saggese and Yongge Wang for their collaboration and insightful contributions to this work.}


\end{frontmatter}



\section{Introduction}

The rapid advancement of quantum computing poses a growing challenge to contemporary
cryptographic infrastructure. Algorithms that underpin Internet security, digital identity, and
data protection, most notably RSA, ECC, and Diffie–Hellman are expected to become vulnerable
once large-scale quantum computers are realized. While PQ cryptography standards
are emerging through initiatives led by NIST \cite{nistir8105} and ETSI \cite{ETSI_TR_103_619_2020}, most enterprises today lack the analytical
frameworks needed to quantify their exposure or track progress toward quantum-safe migration \cite{Le2025CybersecurityAnalytics}.
Existing approaches to cybersecurity risk scoring and maturity assessment remain largely
pre-quantum and fail to capture the systemic dependencies between assets, protocols, and
cryptographic domains that determine enterprise-level vulnerability.

This paper introduces an end-to-end approach to calculating enterprise PQ readiness score, using a knowledge graph based methodology for structuring the data. The framework represents assets and their dependencies as a heterogeneous graph
$G=(V,E)$, where vertices denote digital entities, such as servers, keys, certificates, or services and edges capture cryptographic or logical relationships. Each node is annotated with
PQ resistance and business weight allowing dynamic computation of risk propagation through both internal and external dependencies.

The knowledge graph based framework for PQ cybersecurity assessment uses large language models (LLMs) with human-in-the-loop (HiL) validation for analytics and operational calculations \cite{Tsaneva2025KGValidation}. Enterprise cybersecurity assets, including infrastructure components, cryptographic primitives, certificates, keys, services, and their interdependencies, are represented as a heterogeneous knowledge graph that captures how cryptographic weaknesses propagate across technical and organizational boundaries. LLMs are used in two complementary roles within the framework. First, they support scalable validation of graph entities and relationships through hybrid LLM and HiL workflows that combine automated validation with expert oversight. Second, they are used to infer and assign semantic and contextual weights to assets and dependencies, which directly inform downstream PQ exposure and readiness scoring. By embedding LLM-assisted validation and weighting into the graph construction and scoring pipeline, the proposed approach enables explainable, continuously updatable, and enterprise-scale quantification of PQ risk, bridging cryptographic telemetry with decision-relevant PQ readiness metrics.

\section{Literature Review}

Recent research underscores the growing need for quantitative and scalable methods to assess
enterprise cybersecurity posture. \cite{Cremer2022CyberRisk} provide a comprehensive review of cyber-risk and
cybersecurity data availability, highlighting the lack of standardized, interoperable datasets
for modeling organizational exposure. \cite{Zadeh2023Cybersecurity} propose a quantitative risk quantification and
classification framework that links breach severity with control effectiveness, offering a
decision-analytic foundation for enterprise-level risk scoring. \cite{Le2025CybersecurityAnalytics} conduct a systematic literature review of
cybersecurity analytics in enterprise environments, emphasizing automation, data fusion,
and advanced reasoning capabilities that directly inform our knowledge graph
architecture for scalable PQ risk evaluation. \cite{Erlemann2022CVMChangePoints} studied nonparametric change-point detection methods that identify abrupt distributional shifts, a principle underlying anomaly detection techniques widely used in cybersecurity monitoring and intrusion detection.

Knowledge graphs and graph-based reasoning are increasingly used to model complex cyber
dependencies. 
\cite{Bolton2023CybersecurityKG} demonstrate how cybersecurity knowledge graphs
map to MITRE ATT\&CK domains to enable structured reasoning across assets, vulnerabilities,
and attack vectors.
Similarly, \cite{Sikos2023CybersecurityKG} formalizes a semantic framework for
cybersecurity knowledge graphs, outlining how ontological relationships and graph analytics
can improve threat detection and explainability.
These works collectively establish the foundation for representing enterprise cryptographic
dependencies as a graph structure.

Parallel efforts in cyber-risk classification and maturity assessment have aimed to translate
qualitative posture evaluations into quantifiable scores. \cite{Bernardo2025CybersecurityMaturity} introduce a NIST-CSF–aligned maturity
assessment model that operationalizes cybersecurity evaluation through a dual-survey approach,
producing standardized posture indices across domains. \cite{Zou2019AdversarialDefenseSurvey, wang2008probabilisticAG, zenitani2023attackGraphGuide, chen2024bayesianAGpower} review adversarial machine learning methods
and highlight the need for interpretability and resilience in security analytics, an insight
that motivates the use of Shapley-value attribution within PQ-Score to provide transparent,
domain-level explanations of PQ exposure.

Finally, \cite{Wang2012PKCS} offers a detailed overview of public key cryptography standards, which remain foundational to most enterprise encryption infrastructures.
These standards form the baseline from which this paper evaluates the transition toward
PQ readiness.

\section{Methodology}
\subsection{External Asset Discovery}

External asset discovery provides comprehensive visibility into an organization's internet-facing attack surface, a critical foundation for cybersecurity analytics \cite{Swetha2024ComparativeVulnTools}. However, the heterogeneous nature of discovery data sources (DNS records, port scans, TLS certificates, vulnerability databases) and the complex relationships between discovered entities present significant integration challenges \cite{kott2013cmrs}. Traditional relational approaches struggle to represent the multi-dimensional relationships between assets, vulnerabilities, and attack vectors. Our implementation addresses these challenges through a graph-based knowledge representation architecture that enables both automated reasoning and interactive exploration of the attack surface \citep{Sikos2023CybersecurityKG, Bolton2023CybersecurityKG}.

\begin{figure}[H]
\centering
\vspace{-5mm} 
\includegraphics[width=\textwidth]{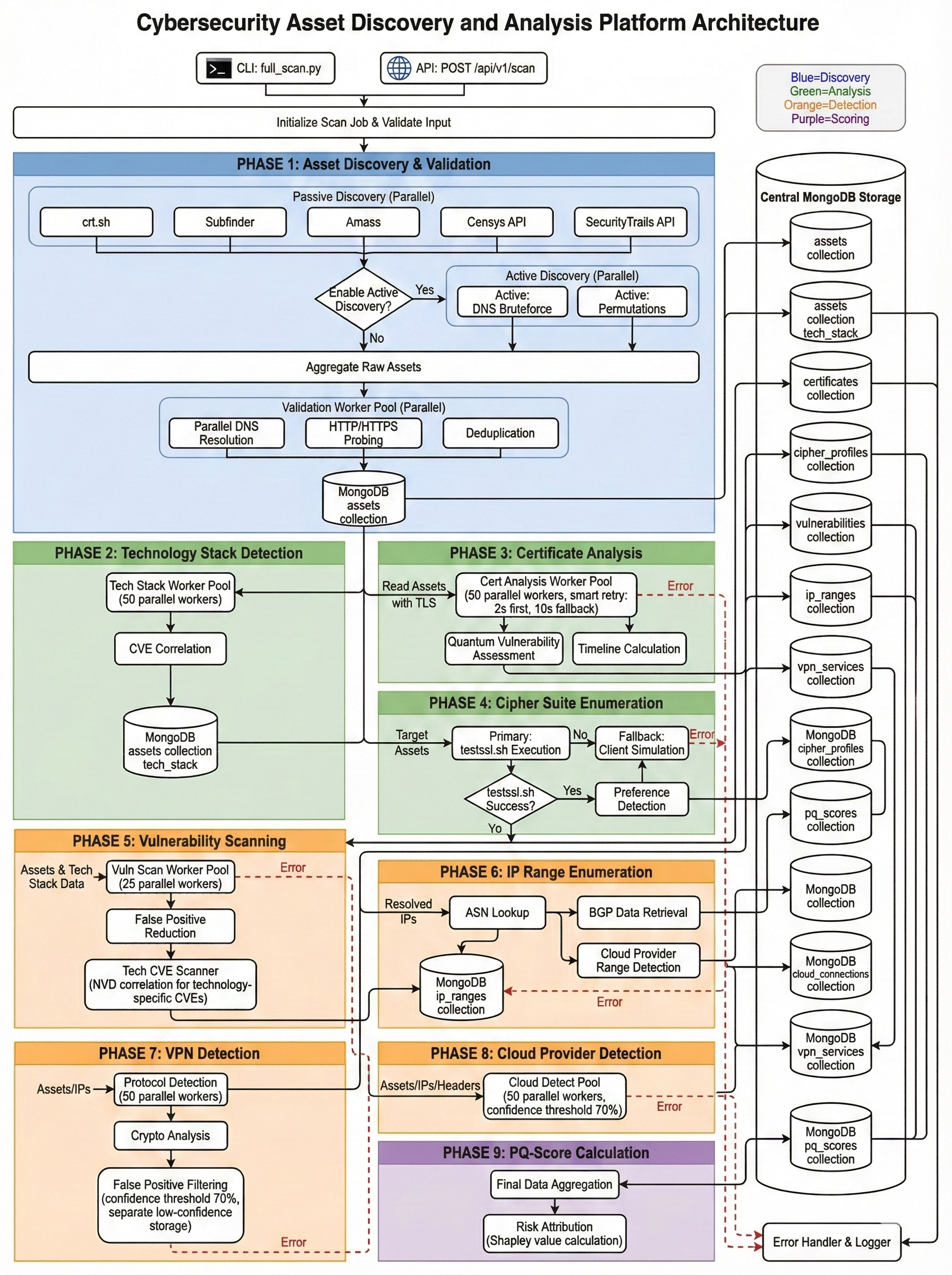}
\caption{External Asset Discovery System Architecture. The scanner framework orchestrates $18$ specialized Python modules across six primary risk categories. ThreadPoolExecutor enables parallel execution with configurable worker pools (50 concurrent workers for certificate extraction, 25 for vulnerability scanning). Real-time API integration retrieves threat intelligence from multiple authoritative sources including NIST NVD, CISA KEV, and EPSS. The ETL pipeline transforms scan results before MongoDB ingestion. Knowledge graph construction processes scanner outputs through GraphBuilderService with LLM-assisted validation.}
\label{fig:scanner_dataflow}
\end{figure}

\textbf{Graph-Based Discovery Architecture.}
The system employs $18$ specialized scanner modules organized by risk factors and asset types, executing in parallel to gather data across multiple dimensions: DNS infrastructure (A/AAAA/MX/TXT/NS records with DNSSEC validation), subdomain enumeration (Certificate Transparency logs and active DNS queries), network services (nmap-based port scanning with CPE matching), and cryptographic implementations (TLS/SSL and SSH analysis including PQ vulnerability assessment). A message queue architecture (Figure \ref{fig:scanner_dataflow}) decouples discovery operations from processing, enabling horizontal scaling and prioritization of urgent scans such as zero-day validation. Scanner execution is triggered by web dashboard requests, scheduled jobs, or threat intelligence updates, with results persisted to PostgreSQL before graph transformation.

\textbf{Knowledge Graph Construction and Integration.}
Raw discovery data undergoes ETL transformation to construct graph relationships following our cybersecurity ontology \cite{zhao2024surveyCKG}. The pipeline: (1) deduplicates assets discovered by multiple scanners, (2) resolves entity identities across data sources (e.g., linking IP addresses to domains and autonomous systems), (3) enriches with threat intelligence correlation from eight vulnerability sources (CISA KEV, NIST NVD, CVE-Search/CIRCL, EPSS, OSV, GitHub Security Advisories, Vulners, and local CVE database), and (4) constructs typed relationships in Neo4j representing dependencies, exposures, and potential attack paths. Vulnerability sources are weighted by reliability (CISA KEV: 1.0, NIST NVD: 0.9, CVE-Search: 0.8) with concurrent queries, timeout controls, and automatic deduplication \citep{Zadeh2023Cybersecurity}. CVSS scores are normalized across versions (v2/v3.0/v3.1) for consistent severity assessment.

The resulting graph structure enables complex queries that traverse relationships to identify attack paths \cite{Kordy2014AttackDefenseTrees, munoz2017efficientAG, alhomidi2014agrisk}, certificate dependencies, and technology stack vulnerabilities, supporting both automated reasoning and interactive dashboard visualization. For example, analysts can query: "Find all internet-facing servers running vulnerable OpenSSL versions that could enable remote code execution and have privileged access to internal databases," traversing service→software→vulnerability→database relationships that would require multiple table joins in traditional architectures. This graph-based approach facilitates the attack scenario modeling and threat intelligence correlation described in subsequent sections, while providing the semantic foundation for machine learning-based anomaly detection.

\textbf{Security and Compliance.}
The discovery infrastructure implements defense-in-depth controls aligned with NIST Cybersecurity Framework requirements \citep{Bernardo2025CybersecurityMaturity}: containerized scanner execution with resource quotas, TLS 1.3 encryption in transit, AES-256 encryption at rest on FIPS 140-2 validated storage, role-based access control for scan invocation, and comprehensive audit logging. The system excludes PII from collection scope to maintain privacy-by-design principles and GDPR compliance.

\subsection{Knowledge Graph Design}
The knowledge graph is designed as a heterogeneous, property-graph representation of the enterprise cryptographic landscape, integrating both externally discovered assets and internally curated configuration data into a single analytical model. Nodes represent infrastructure assets (e.g., hosts, services, certificates, keys, identities, data stores) as well as more abstract organizational concepts (e.g., business units, cryptographic domains, regulatory obligations), while edges encode directed dependencies that capture how compromise or cryptographic weakness can propagate through the environment.

At the core of the model is the asset layer. Each compute or network node (e.g., server, virtual machine, container, network device) is represented as an Asset node with attributes (e.g., asset\_id, platform, ip\_address, criticality\_weight $w_i)$, resistance $R_i$, etc). Attributes are selected first by defining what decisions the graph must support. For PQ-exposure analysis, we ask:
\begin{itemize}
\item Does this attribute help quantify quantum-related risk or business impact?
\item Does it support propagation of risk across dependencies?
\item Does it help prioritize remediation efforts?
\end{itemize}
In addition to Asset nodes, the knowledge graph contains specialized node classes, such as service nodes, certificate nodes, key nodes, vulnerability nodes, etc. The decision to represent information as a separate node versus an attribute of an existing node depends on the role that the information plays in reasoning, traversal, and inference within the graph. A concept is modeled as a new node when it represents a distinct real-world entity with its own lifecycle, relationships, and potential to influence or propagate exposure across the system. For example, certificates, cryptographic keys, or vulnerabilities are created as independent nodes because they are shared across multiple assets or services, change over time, and participate in many-to-many relationships. Representing them as nodes allows the graph to quantify the cascading impact of weaknesses and provide explainable remediation paths.

\begin{figure}[H]
\centering
\includegraphics[width=\textwidth]{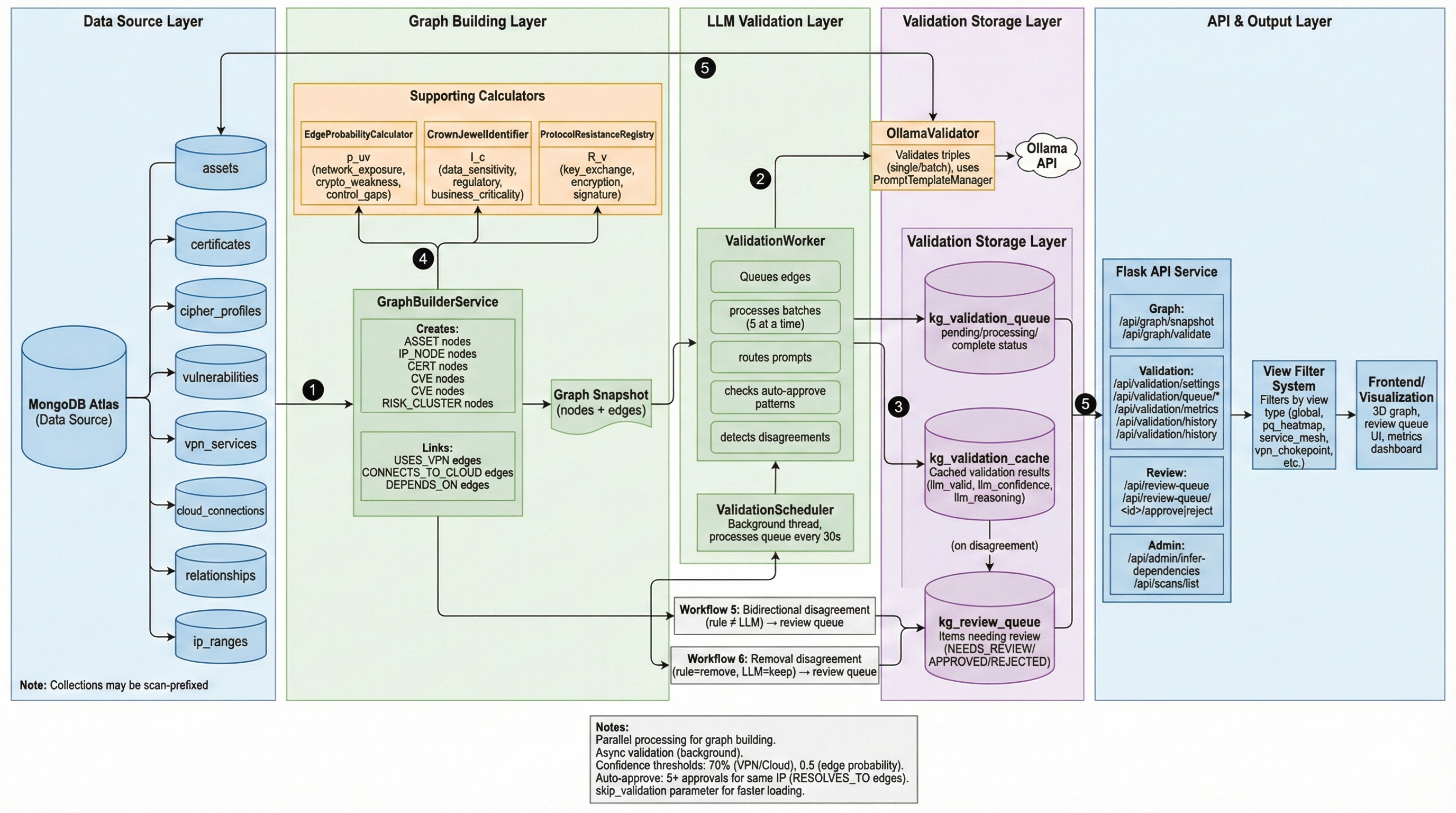}
\caption{Knowledge Graph Construction and LLM-Assisted Validation Architecture. The system integrates five functional layers: (1) Data Source Layer containing MongoDB Atlas collections for assets, certificates, cryptographic profiles, vulnerabilities, and network services; (2) Graph Building Layer orchestrating the GraphBuilderService, which creates typed nodes (ASSET, IP\_NODE, CERT\_NODE, CVE, RISK\_CLUSTER) and relationships (USES, CONNECTS\_TO, EXPOSES, DEPENDS\_ON) with support from EdgeProbabilityCalculator, CrownJewelIdentifier, and ProtocolResistanceRegistry; (3) LLM Validation Layer deploying OllamaValidator for batch processing and ValidationWorker for queue management with automated approval thresholds and disagreement detection; (4) Validation Storage Layer maintaining three MongoDB collections (kg\_validation\_queue, kg\_validation\_cache, kg\_review\_queue) with bidirectional disagreement workflows (Workflow 5: rule $\neq$ LLM triggers review; Workflow 6: rule=remove, LLM=keep triggers review); and (5) API \& Output Layer providing Flask endpoints for graph snapshots, validation settings, and review queue management with frontend visualization through filtered views by validation status, PQ heatmap, service mesh, and VPN chokepoint analysis. ValidationScheduler operates asynchronously with 30-second intervals to process pending validations. Auto-approve patterns expedite high-confidence edges (e.g., RESOLVES\_TO edges with probability $>0.5$), while disagreement detection routes contested edges to human oversight. Specialized validation prompts are tailored to ten distinct relationship semantics, including USES edges (asset-certificate authentication), CONNECTS\_TO edges (network connectivity plausibility), EXPOSES edges (vulnerability applicability), and DEPENDS\_ON edges (cloud/VPN dependencies), with each prompt incorporating post-quantum awareness for cryptographic algorithm evaluation.}
\label{fig:kg_llm_architecture}
\end{figure}

Edges in the knowledge graph represent the directional relationships and dependencies through which cryptographic posture, operational risk, and business impact propagate across the enterprise system, as previously considered by \cite{singhal2011probabilistic}. Unlike attributes, which provide descriptive context to individual nodes, edges define how entities interact with one another. For example, an edge connects a certificate node to a data storage node, representing that the certificate is used to encrypt traffic or protect data access for that repository.

\subsection{LLM-Assisted Knowledge Graph Validation}

The knowledge graph validation framework employs a two-stage methodology that integrates large language models (LLMs) with human-in-the-loop (HiL) oversight to ensure data quality while maintaining scalability. This approach extends recent advances in hybrid AI-human workflows for scientific knowledge graph validation \cite{Tsaneva2024LLMValidation}, adapting their methodology for cybersecurity domain-specific requirements and PQ risk assessment.

\subsubsection{Stage 1: Asset and Relationship Validation}

The first validation stage verifies the existence and correctness of discovered assets (graph nodes) and their cryptographic or operational dependencies (graph edges). Scanner-discovered entities undergo automated classification using large language models to determine validity. Adapting the validation architecture from prior work on knowledge graph validation tasks \cite{Tsaneva2024LLMValidation}, our implementation leverages local LLM inference to assess whether discovered relationships represent authentic enterprise infrastructure while maintaining data privacy through air-gapped deployment.

For each triple $(s, p, o)$ representing a relationship between source asset $s$, predicate $p$, and target asset $o$, the LLM validator processes batches of candidate edges through the \texttt{OllamaValidator} component. The validator receives structured prompts tailored to relationship semantics. For \texttt{USES} edges connecting assets to certificates, the prompt framework is:

\begin{quote}
\small
\textit{You are a cybersecurity expert analyzing infrastructure dependencies. Evaluate whether this relationship is technically valid: [Asset FQDN: \{source\}] USES [Certificate fingerprint: \{cert\_sha256\}, algorithm: \{algorithm\}, key\_size: \{key\_size\}].}

\textit{Consider: (1) Does this asset type typically use X.509 certificates? (2) Is the certificate algorithm appropriate for the asset's service type? (3) Is the relationship structurally plausible in enterprise networks?}

\textit{Respond with valid=true/false and confidence (0-1).}
\end{quote}

Each validation batch undergoes processing with majority voting to reduce stochastic variability. Following validation, human-in-the-loop review is triggered for edges where LLM confidence falls below the disagreement threshold (0.7 for \texttt{VPN\_SERVICE} and \texttt{CLOUD\_CONNECTION} relationships, 0.5 for general edge probability assessments). Enterprise cybersecurity teams perform validation through a web-based review queue interface, examining asset classifications and confirming or rejecting automated decisions. This selective human oversight implements the disagreement-strategy approach where minimal manual effort (validating $0.4\%$ of generated edges in our implementation) maintains precision while achieving high validity rates. In our production deployment, the system validated nearly $4000$ relationships with a $99.6\%$ validity rate, $94.6\%$ average confidence, and $0.4\%$ disagreement rate requiring human review.

\subsubsection{Stage 2: Attribute Validation and Risk Scoring}

The second validation stage assigns quantitative risk attributes essential for PQ exposure calculation. For each validated node $v$, the LLM estimates PQ resistance score $R_v \in [0,1]$ and business weight $w_v \in [0,1]$. For validated edges $(u,v)$, the system assigns exploitability weight $p_{uv} \in [0,1]$ based on control effectiveness, privilege requirements, and network exposure.

Resistance score assignment leverages the \texttt{ProtocolResistanceRegistry}, which maintains mappings from cryptographic protocols to quantum vulnerability assessments. The registry classifies algorithms into three categories: vulnerable (e.g., RSA-2048, ECDSA-P256: $R_v \in [0, 0.3]$), transitional (e.g., RSA-4096: $R_v \in [0.3, 0.7]$), and quantum-safe (e.g., CRYSTALS-Kyber, CRYSTALS-Dilithium: $R_v \in [0.9, 1.0]$). For complex assets with multiple cryptographic implementations, the weakest-link principle applies: overall asset resistance equals the minimum resistance across all deployed algorithms.

Business weight estimation ($w_v$) incorporates contextual information including service criticality, data sensitivity classifications, and organizational impact. The LLM receives asset profiles containing discovered services, network exposure indicators, and vulnerability severity distributions. For crown jewel identification, the prompt framework follows:

\begin{quote}
\small
\textit{Analyze this enterprise asset: [FQDN: \{fqdn\}, Services: \{service\_list\}, Vulnerabilities: \{vuln\_summary\}, Network exposure: \{exposure\_level\}].}

\textit{Assign business weight (0-1): 0.9-1.0 = mission-critical (authentication servers, databases with PII); 0.6-0.8 = important operational systems; 0.3-0.5 = standard infrastructure; 0.0-0.2 = development/test environments.}

\textit{Consider: data sensitivity, service criticality, regulatory compliance requirements, operational dependencies.}
\end{quote}

Edge exploitability weights ($p_{uv}$) quantify the probability that an adversary successfully traverses a dependency relationship. Three supporting calculators inform edge probability estimation:

\begin{itemize}[noitemsep]
    \item \textbf{EdgeProbabilityCalculator}: Evaluates network exposure, cryptographic weaknesses, and control gaps to produce base exploitability scores.
    \item \textbf{CrownJewelIdentifier}: Assesses data sensitivity, regulatory obligations, and business criticality to determine asset importance $I_c$.
    \item \textbf{ProtocolResistanceRegistry}: Maps cryptographic protocols to PQ resistance scores based on key exchange algorithms, signature schemes, and encryption primitives.
\end{itemize}

\subsubsection{Validation Workflows and Human Oversight}

The system implements bidirectional disagreement workflows that balance automated throughput with human quality assurance. When the LLM validator produces results contradicting scanner findings or rule-based heuristics, edges are routed to the \texttt{kg\_review\_queue} collection for manual inspection. Conversely, when multiple validators agree and confidence exceeds auto-approval thresholds, edges are immediately incorporated into the knowledge graph without human review.

This architecture follows agreement-based validation strategies. The disagreement detection mechanism examines three conditions: (1) LLM confidence below threshold; (2) LLM decision conflicts with rule-based validator; (3) asset is classified as high-criticality ($w_v > 0.8$) or crown jewel ($I_c > 0.9$).

Validation state management utilizes three MongoDB collections. The \texttt{kg\_validation\_queue} collection tracks pending validations with processing status indicators (\texttt{pending}, \texttt{processing}, \texttt{complete}).
\texttt{kg\_validation\_cache} collection persists LLM reasoning, confidence scores, and validation outcomes to prevent duplicate evaluations. The \texttt{kg\_review\_queue} collection maintains human review items, storing disagreement reasons, original LLM responses, and cybersecurity team decisions. A background \texttt{ValidationScheduler} processes the validation queue every 30 seconds, enabling continuous validation of newly discovered assets while maintaining comprehensive audit trails for regulatory compliance.

\subsubsection{Validation Prompts for Edge Types}

The validation system employs ten specialized prompts tailored to distinct relationship semantics. Representative examples include:

\textbf{USES Edge Validation (Asset $\rightarrow$ Certificate):}
Assesses whether an asset legitimately relies on a specific certificate for secure communications, considering algorithm appropriateness and enterprise deployment patterns.

\textbf{CONNECTS\_TO Edge Validation (Asset $\rightarrow$ Asset):}
Evaluates network connectivity plausibility based on service types, port configurations, and typical enterprise architectures.

\textbf{EXPOSES Edge Validation (Asset $\rightarrow$ CVE):}
Verifies vulnerability applicability by matching asset technology stacks to affected software versions and CVE applicability statements.

\textbf{DEPENDS\_ON Edge Validation (Asset $\rightarrow$ Cloud/VPN):}
Analyzes whether remote access or cloud provider dependencies align with observed network configurations and service requirements.

Each prompt incorporates PQ awareness by requesting explicit consideration of cryptographic algorithm choices, key sizes, and quantum vulnerability timelines. This domain specialization distinguishes our validation approach from general-purpose knowledge graph validation frameworks.

\subsection{Large Language Models for Validation}

Table~\ref{tab:llm_models} summarizes the large language models employed in the knowledge graph validation pipeline. Due to data sensitivity requirements and the need for air-gapped deployment in enterprise cybersecurity environments, we utilize the Ollama framework as a local LLM inference server. Our implementation was tested with Google's Gemma 3:4b and Gemma 3:12b models, with the 12b variant selected for production deployment based on superior accuracy and reduced false positive rates. Initial testing with the 4b model revealed validation inconsistencies that were eliminated when migrating to the 12b configuration. We additionally evaluated the 27b model but observed no accuracy improvements over 12b while experiencing significantly longer per-node processing times, making the 12b model optimal for our real-time validation requirements.

\begin{table}[htbp]
\centering
\small
\begin{tabular}{llll}
\toprule
\textbf{Model} & \textbf{Version/Identifier} & \textbf{Provider} & \textbf{Deployment Mode} \\
\midrule
Gemma 3 & \texttt{gemma3:12b} & Google/Ollama & Local Inference \\
\bottomrule
\end{tabular}
\caption{Large language models used for knowledge graph validation. Gemma 3:12b serves as the production validator for asset classification, relationship verification, and quantitative risk attribute assignment ($R_v$, $w_v$, $p_{uv}$). Ollama provides local inference capabilities for air-gapped enterprise deployments and scenarios requiring on-premises processing of sensitive infrastructure data. Model selection was based on empirical testing comparing Gemma 3:4b, 3:12b, and 3:27b variants.}
\label{tab:llm_models}
\end{table}

The \texttt{OllamaValidator} component interfaces with the local Ollama service endpoint (\texttt{http://localhost:11434}) through a unified API abstraction, using standardized prompt structures that ensure validation consistency across different model configurations. Configuration parameters including model selection, batch processing parameters, and confidence thresholds are stored in MongoDB's \texttt{validation\_settings} collection, enabling dynamic adjustment without service restarts.

\subsection{Post-Quantum Readiness Scoring Framework}

Building upon established risk assessment methodologies \cite{baseri2025evaluation}, we 
extend the analysis of PQ cryptographic vulnerabilities by modeling enterprise 
assets as a knowledge graph where Shapley values attribute exposure across cryptographic 
domains. While \cite{baseri2025evaluation} focus on pre-migration, through-migration, and 
post-migration threat analysis using STRIDE, our framework quantifies the marginal 
contribution of each domain to overall quantum risk.

To attribute PQ vulnerability across cryptographic domains in a 
complex enterprise system, we model all digital assets and their dependencies as a 
knowledge graph $G=(V,E)$, where vertices $v \in V$ represent assets and edges $(u \!\to\! v) \in E$ encode 
cryptographic or logical dependencies between them \cite{Sikos2023CybersecurityKG}. Each node is annotated with 
attributes $\{R_v, w_v\}$ denoting, respectively, its PQ resistance 
score and business or operational weight.
Edges are assigned exploitability weights $p_{uv} \in [0,1]$ derived from control 
gaps or network exposure.  Let $A \subseteq V$ denote potential adversarial 
entry points, and $C \subseteq V$ denote ``crown-jewel'' assets with normalized 
impact $I_c \in [0,1]$. For example, in practice, $C$ may represent a set of databases containing personally identifiable information.

\subsection{PQ Scoring Algorithm}
The base PQ exposure functional $E(S)$ quantifies the expected risk of compromise 
given a subset of domains $S$ under analysis. Two complementary formulations are 
used depending on graph size. First, there is exact path-based model $E_E$ for smaller graphs consisting of hundreds of nodes. Prior work has derived vulnerability metrics over attack graphs \cite{gueye2012vulnerability}.
    \begin{equation}
        E_E(S) = 
        \frac{\sum_{c \in C} I_c 
            \Big( 1 - 
                \prod_{\pi \in \Pi_S(A \rightsquigarrow c)} 
                \big(1 - P_S(\pi)\big)
            \Big)
        }{\sum_{c \in C} I_c},
        \label{eq:path-exposure}
    \end{equation}
    where $\Pi_S(A \rightsquigarrow c)$ is the set of minimal simple paths 
    connecting entry points $A$ to target $c$ \cite{Sheyner2002Automated}.
    Each path probability $P_S(\pi)$ captures both cryptographic weakness and 
    dependency exploitability:
    \begin{equation}
        P_S(\pi) = 
        \Bigg( \prod_{(u \to v) \in \pi} p_{uv} \Bigg)
        \Bigg( \prod_{v \in \pi} (1 - R_v) \Bigg)
        \chi_S(\pi),
    \end{equation}
    with $\chi_S(\pi) \in [0,1]$ representing the fraction of nodes or edges in 
    $\pi$ that belong to the coalition of domains $S$.
    
  Secondly, there is the all-walks (Katz) approximation $E_A$ for large graphs \cite{katz1953,Beres2018TemporalKatz}. Following the classical Katz centrality framework \cite{katz1953}, this involves a matrix-based approximation which captures all attenuated paths \cite{vandergrinten2018katz}.
\begin{equation}
    E_A(S) = 
     a^{\top} (I - \alpha W_S)^{-1} \, b,
    \label{eq:katz-exposure}
\end{equation}
where $W_{uv} = p_{uv}(1-R_v)w_v m_{v,S}$ is the weighted adjacency matrix, $m_{v,S} \in \{0,1\}$ is an indicator function equal to $1$ if node $v$ belongs to the domain coalition $S$ and $0$ otherwise, 
$\alpha < 1/\rho(W_S)$ is the attenuation factor ensuring convergence (where $\rho(W_S)$ denotes the spectral radius), $a$ is the entry-node indicator vector representing potential adversarial access points, 
and $b$ is the target vector of crown-jewel impacts $I_c$.
This formulation efficiently aggregates all dependency walks while exponentially down-weighting longer paths via the attenuation parameter $\alpha$.

To allocate overall PQ exposure fairly among interacting cryptographic domains, like transport, data-at-rest, identity, code-signing, 
we compute the Shapley value $\phi_d$ for each domain $d \in D$ and the formula is a Theorem from \cite{shapley1953value}.
\begin{equation}
    \phi_d =
    \sum_{S \subseteq D \setminus \{d\}}
    \frac{|S|!\,(|D|-|S|-1)!}{|D|!}
    \big( E(S \cup \{d\}) - E(S) \big),
    \label{eq:shapley}
\end{equation}
where $E(S)$ is either $E_E(S)$ or $E_A(S)$.  The Shapley value $\phi_d$ represents the marginal 
contribution of domain $d$ to the total PQ exposure, averaged over all possible 
orderings of domain inclusion.  These values satisfy $\sum_{d \in D} \phi_d = E(D)$ 
and yield a principled decomposition of enterprise-level PQ risk across overlapping 
functional areas.

Shapley values provide a principled way to attribute each asset's contribution to an overall enterprise risk or exposure score. 
Consider a PQ readiness model that predicts an organization’s total cryptographic vulnerability based on three features: key length, algorithm type, and network exposure. 
If removing network exposure from the model substantially lowers the predicted risk, while removing key length or algorithm type has smaller marginal effects, the Shapley framework quantifies each feature’s average contribution across all possible combinations of features. 
For example, network exposure may account for 0.45 of the overall vulnerability score, algorithm type for 0.35, and key length for 0.20, ensuring a fair, additive explanation of how individual asset attributes drive the enterprise’s aggregate PQ exposure.

\subsection{PQ Score Normalization and Validation}
To ensure comparability and interpretability of PQ exposure values across domains and graph scales, all exposure scores are normalized and empirically validated. The raw exposure functions $E(S)$ from Equations~\ref{eq:path-exposure} and~\ref{eq:katz-exposure} yield dimensionless values on $[0,1]$, but their magnitudes depend on heterogeneous asset weights $w_v$, resistance scores $R_v$, and impact factors $I_c$. We therefore define a normalized exposure
\begin{equation}
    \widehat{E}(S) = \frac{E(S)}{E(S_{\max})},
    \label{eq:normalized}
\end{equation}
The normalization constant $E(S_{\max})$ represents the hypothetical maximum exposure obtained by setting all resistance scores $R_v = 0$ (complete quantum vulnerability) while preserving the original network structure and dependencies. The normalized score $\widehat{E}(S) \in [0,1]$ quantifies the proportion of theoretical maximum risk still present under current PQ-readiness conditions. Domain-level Shapley attributions $\phi_d$ are normalized such that $\sum_{d \in D} \phi_d = \widehat{E}(D)$, enabling fair cross-domain aggregation and longitudinal comparison. Temporal changes in enterprise readiness are evaluated by the relative improvement $\Delta E_t = (\widehat{E}_t - \widehat{E}_{t_0}) / \widehat{E}_{t_0}$, with negative values indicating successful mitigation. Model validation combines correlation, sensitivity, and empirical benchmarking: correlation between the path-based and all-walks formulations confirms internal consistency; sensitivity analysis of $\partial E / \partial R_v$, $\partial E / \partial p_{uv}$, and $\partial E / \partial I_c$ verifies monotonic response to security improvements; and simulation-based benchmarking aligns PQ exposure predictions with observed compromise probabilities in synthetic harvest-now–decrypt-later experiments. Finally, a composite Post-Quantum Readiness Index is defined as
\begin{equation}
    \mathrm{PQRI} = 100 \times (1 - \widehat{E}(D)),
\end{equation}
providing an intuitive scale where higher values indicate stronger organizational resilience to quantum-enabled cryptographic compromise.

\section{Conclusion}

This paper presented a knowledge graph based framework for quantifying enterprise PQ readiness by modeling cryptographic assets, dependencies, and risk propagation across complex infrastructures. By representing enterprise environments as heterogeneous graphs, the approach captures systemic interactions that are not addressed by asset-centric or checklist-based assessments, and enables explainable attribution of PQ exposure across cryptographic domains using graph-theoretic risk functionals and Shapley value decomposition. The integration of large language models with human-in-the-loop validation supports scalable asset classification and relationship verification while preserving expert oversight for high-impact decisions. The resulting normalized readiness metrics are suitable for continuous monitoring, comparative analysis, and remediation prioritization, providing decision-makers with actionable insight into migration progress. Although detailed deployment results cannot be publicly disclosed due to security considerations, internal validation demonstrates the practical feasibility of the framework on real-world enterprise infrastructure, establishing a principled and extensible foundation for enterprise-scale PQ risk assessment.

\section*{Declaration of Competing Interest}

The authors declare that they have no known competing financial interests or personal relationships that could have appeared to influence the work reported in this paper.

\section*{Acknowledgments}

This research was supported in part by the NSF I-Corps program. 
The work also aligns with ongoing research activities under NSF 25-515 
(Security, Privacy, and Trust in Cyberspace).

\section*{Declaration of Generative AI and AI-assisted Technologies in the Writing Process}

During the preparation of this work, the authors used generative AI tools (ChatGPT 5.2 and Opus 4.5) to assist with improving the readability and language of the manuscript and to support code development tasks such as suggesting code structure, refactoring patterns, and documentation. All AI-assisted outputs were carefully reviewed, validated, and edited by the authors. The authors take full responsibility for the accuracy, integrity, and originality of the manuscript and all associated software artifacts.

\bibliographystyle{elsarticle-num} 
\bibliography{ref}
\end{document}